\documentclass[10.5pt]{article}

\def\lromn#1{\uppercase\expandafter{\romannumeral#1}}

\def\lromn#1{\uppercase\expandafter{\romannumeral#1}}

\usepackage[pdftex]{graphicx}

\begin{document}

\vspace{2cm}
\begin{center}
\begin{Large}

{\bf Experimental method of detecting relic neutrino by atomic de-excitation}
\end{Large}

\vspace{2cm}
M.~Yoshimura, N. Sasao$^{\dagger}$, and M. Tanaka$^{\ddagger}$

\vspace{0.5cm}
Center of Quantum Universe, Faculty of
Science, Okayama University \\
Tsushima-naka 3-1-1 Kita-ku Okayama
700-8530 Japan

$^{\dagger}$
Research Core for Extreme Quantum World,
Okayama University \\
Tsushima-naka 3-1-1 Kita-ku Okayama
700-8530 Japan \\

$^{\ddagger}$
Department of Physics, Graduate School of Science, 
             Osaka University, Toyonaka, Osaka 560-0043, Japan

\end{center}

\vspace{5cm}

\begin{center}
\begin{Large}
{\bf ABSTRACT}
\end{Large}
\end{center}

The cosmic background neutrino
of temperature  1.9 K affects
rates of radiative  emission of neutrino pair (RENP) 
from metastable excited atoms,
since its presence  blocks the pair emission by
the Pauli exclusion principle.
We quantitatively investigate how the Pauli blocking
distorts the photon energy spectrum and
calculate its sensitivity to cosmic parameters
such as the neutrino temperature and its
chemical potential.
Important quantities for high sensitivities to these parameter measurement
are found to be the level spacing of atomic de-excitation
and the unknown mass value of lightest neutrino,
in particular their mutual relation.

\vspace{4cm}
PACS numbers
\hspace{0.5cm} 
13.15.+g, 
14.60.Pq, 
98.80.Es  

Keywords
\hspace{0.5cm} 
relic neutrino,
Pauli exclusion principle,
big bang cosmology,
nucleo-synthesis,
neutrino mass,
cosmological lepton asymmetry

\newpage
The relic cosmic neutrino of temperature 
$(4/11)^{1/3} T_{\gamma} \sim $1.9K (with $T_{\gamma}$
the cosmic microwave temperature)
is undoubtedly one of the most important predictions
of the big bang cosmology \cite{cosmology}.
Detection of relic neutrino would give a strong support for
nucleo-synthesis theory that explains the origin of cosmic
light elements such as $^4$He.
Various ideas of  experimental method of 
relic neutrino detection have been discussed in the literature
\cite{zt beta, relic old, stodolsky, Z production via relic, {irvine}, my 07, raghavan, mangano, relic review, ptolemy}.

In the present work
we  propose a new experimental method using excited
atomic targets.
The idea is based on the fact that radiative emission of neutrino pair (RENP)
\cite{renp overview}
is affected by the Pauli blocking of ambient cosmic neutrinos 
\cite{my 07 comment}.
We shall give an answer to the fundamental issue of how sensitive the Pauli blocking 
effect is to
determination of cosmological parameters, the neutrino temperature
and the chemical potential
which is related to the lepton asymmetry of our universe.

The process we use is atomic de-excitation from
a metastable state 
$|e\rangle$; $|e\rangle \to | g\rangle + \gamma + \nu_i \bar{\nu}_j$
(anti-neutrino $\bar{\nu}_j$ is identical to $ \nu_j$ in the case of Majorana neutrino).
Energy spectrum of the photon $\gamma$ and parity violating quantities
such as the asymmetry of rates under the
magnetic field reversal \cite{renp pv} are measured in RENP.
$\nu_i (i=1,2,3)$ is a mass eigenstate of neutrinos and a mixture of 
neutrino species 
$\nu_e, \nu_{\mu}, \nu_{\tau}$
that appear in the weak decay of elementary particles.
Neutrino oscillation experiments \cite{neutrino oscillation} have determined 
two mass squared differences, $( \sim 50 {\rm meV})^2$ and $ (\sim 10 {\rm meV})^2$,
and three mixing angles in a theoretical framework of an extended standard
gauge theory where finite neutrino masses and $3\times 3$ unitary mixing
are introduced as an extra assumption.
RENP process predicts a continuous photon energy spectrum at 
$\omega < \omega_{ij}$
(we use the natural unit of $\hbar = c = k_B = 1$ such that 
$\omega$ is the photon energy).
Six thresholds  are given by
$\omega_{ij} =\epsilon_{eg}/2 - (m_i+m_j)^2/(2 \epsilon_{eg}) $
with $\epsilon_{eg}$ the level spacing of excitation.
Since RENP occurs via stimulated photon emission by
trigger lasers, 
decomposition into neutrino mass eigen-states is made possible
by the excellent resolution of trigger laser frequencies.
RENP experimental project \cite{renp overview} has been proposed to determine 
the smallest neutrino mass $m_0$,
to distinguish the Majorana neutrino from the Dirac neutrino and to determine
remaining elements of the mixing matrix, CP violating (CPV) phases including the ones
intrinsic to the Majorana neutrino.

Under the ambient relic neutrino background
RENP rates are reduced by the product of Pauli blocking factors $(1-f_i) (1-\bar{f}_j)$
where $f_i, \bar{f}_j$ are the momentum distribution functions for  mass eigen-states 
$\nu_i, \bar{\nu}_j$.
The Einstein relation in the expanding universe
is $E = \sqrt{p^2+m^2/ (z+1)^2 }, z+1 = a(t)/a(t_d)$ where
$a(t), t_d,  z$ are the cosmic scale factor at the present time,
the decoupling time and the red shift factor
since the neutrino decoupling.
To a good approximation (ignoring the momentum region,
$p < O(100 {\rm meV})/(z+1) \sim O(10^{-11})$ eV of extremely 
small phase space), the neutrino mass term can
be neglected in the distribution functions even at the present epoch.
The  distribution function after the neutrino decoupling
changes under the gravity of the expanding universe
and its present form is given by
\(\:
f_i (p) =1/(e^{p/T_{\nu}- \mu_d/T_d} + 1)
\:\)
where $T_{\nu}$ is the effective neutrino temperature at present
given by $ (4/11)^{1/3}T_{\gamma} \sim 1.9$K.
The quantity related to the chemical potential, $\mu_d/T_d$,
is the ratio of the chemical potential to the temperature
at the epoch of neutrino decoupling \cite{chemical poential}.
For $\bar{\nu}_j$ the chemical potential is sign reversed;
$\mu_d \rightarrow - \mu_d$.
The upper bound 
allowed by nucleo-synthesis is of   $O( 1)$ 
\cite{nucleosynthesis bound}.
Some cosmological models predict a large $\mu_d/T_d$
\cite{model of baryon asymmetry}.

The underlying assumption for description in terms of a single
neutrino temperature $T_{\nu}$
 is that no dramatic entropy generation
occurs at the epoch between decoupling of 
$\nu_{\mu}, \nu_{\tau}$ and electron neutrino $\nu_e$, 
since their decoupling temperatures 
are close; 
$ \sim$ 1.9 MeV for $\nu_e$ decoupling, and $ \sim 3.1$ MeV 
for $\nu_{\mu}, \nu_{\tau}$ decoupling \cite{dolgov}.
The measurement of 1.9 K neutrino temperature different from the
microwave temperature 2.7 K is a clear indication of 
physical process 
that occurred at earlier epochs of a few seconds
after the big bang; electron-positron annihilation.

Effect of the gravitational clustering is expected to be small
in the neutrino mass range of $< O(100)$ meV considered below.
The gravitational clustering of massive neutrinos
enhances distortion of the spectrum further than the case without
the clustering, thus gives
a brighter prospect of relic neutrino detection.
A simple rough estimate of the clustering effect is
to multiply the ratio of the number density of relevant neutrino
in our galaxy to the  cosmic density
$3\zeta(3)T_{\nu}^3/(2\pi^2) \sim 110$cm$^{-3}$.
This ratio may be calculated, if necessary, by solving the gravitational collapse of 
massive, but non-interacting particles under the
gravity of cold dark matter \cite{gravitational clustering},
\cite{gravitational clustering 2}.

Spectrum shape functions previously derived 
without the Pauli blocking effect  \cite{nuclear monopole}, \cite{dpsty-plb} 
are modified by  $(1-f_i) (1-\bar{f}_j)$ for
pair production of  $\nu_i \bar{\nu}_j$ at $\omega < \omega_{ij}$.
The spectral shape function $F^{A}$ 
($A = M $ for the nuclear monopole contribution 
of three thresholds $\omega_{ii}$ \cite{nuclear monopole} 
and $A = S$ for the electron spin contribution 
of much smaller absolute rates \cite{dpsty-plb},
two cases being applicable to atoms of different quantum numbers)
for the neutrino pair production 
of  masses $m_i, m_j$ is calculated as an integral over one of the
neutrino energies;
\begin{eqnarray}
& F_{ij}^{A} (\omega; T_{\nu})=
  \frac{1}{8\pi \omega}\int_{E_-}^{E_+} dE_1\, 
  g_{ij}^A (E_1)\cdot \left(1 - f(\sqrt{E_1^2-m_i^2}) \right)
 \left(1 - \bar{f} (\sqrt{(\epsilon_{eg} - \omega -E_1)^2-m_j^2}) \right)\,,
  \label {threshold kinematical factor}\\
& g_{ii}^M (E) = 
  -E^2 + (\epsilon_{eg} - \omega) E + \frac{1}{2} m_i^2 - \frac{1}{4}
  \epsilon_{eg} (\epsilon_{eg} -2 \omega) +  \delta_M\frac{ m_i^2}{2}\,, \\
& g_{ij}^S(E) = - \frac{1}{3} E^2 + \frac{1}{3} (\epsilon_{eg} - \omega) E
  + \frac{1}{12} \epsilon_{eg} (\epsilon_{eg} -2 \omega) 
  - \frac{1}{12}(m_i^2 + m_j^2) - \delta_M \frac{m_i m_j}{2}\,,\\
& E_{\pm} =
  \frac{1}{2} \left( (\epsilon_{eg} - \omega) (1 +
  \frac{m_i^2 - m_j^2}{\epsilon_{eg}(\epsilon_{eg} - 2\omega)} )
  \pm \omega \Delta_{ij}(\omega)
  \right)\,,\quad
  \Delta_{ij}(\omega) 
  = \left\{
    \left(1 - \frac{ (m_i + m_j)^2}{\epsilon_{eg} (\epsilon_{eg} -2\omega) } 
    \right)
    \left(1 - \frac{ (m_i - m_j)^2}{\epsilon_{eg} (\epsilon_{eg} -2\omega) } 
    \right)
    \right\}^{1/2}\,.
\end{eqnarray}
times factors related to atomic matrix elements
and energy denominators in perturbation theory \cite{renp overview}.
These atomic factors cancel out in 
 the ratio of rates,  the rate with to the rate without the Pauli blocking.
Eq.(\ref{threshold kinematical factor})
is a function of photon energy $\omega$, depending
on five parameters, two cosmological
ones $T_{\nu}, \mu_d/T_d$,  two neutrino masses, $m_i, m_j$,
and the atomic level spacing $\epsilon_{eg}$.
$\delta_M = 1$ for Majorana neutrinos,
arising from the interference term of identical fermions,
and $\delta_M=0$  in its absence for Dirac neutrinos.
In the numerical calculations below we present results for
the Majorana case \cite{dirac}.
One may define the total ratio  adding all pair threshold contributions
 with weights determined by
oscillation data \cite{neutrino oscillation};
$R_A(\omega) =F^A(\omega; T_{\nu})/F^A(\omega; 0)$  
($F^A(\omega; 0)$ is the rate factor without the Pauli blocking).
The theoretically calculated quantity $R_A(\omega)$ shown below is insensitive to relevant transition 
dipole moments and other atomic factors, 
the atomic dependence being essentially given by $\epsilon_{eg}$ alone.
Corresponding experimental values $R_A(\omega)$ need input of
theoretical calculation of rates without the Pauli blocking, which
requires other atomic parameters than $\epsilon_{eg}$.

\begin{figure}
 \centering
\includegraphics[width=26em]{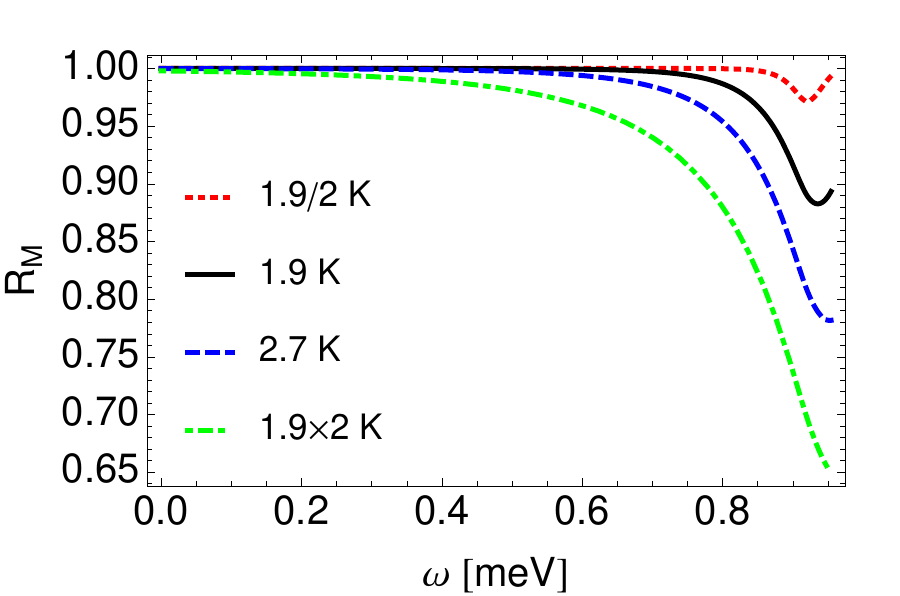}
 \caption{\label {pauli 4ks 11}   
Spectral distortion $R_M(\omega)$ caused by the Pauli blocking of
relic neutrinos,
$T_{\nu}= 1.9/2$ K in  dotted red, 1.9 K in solid black,
2.7 K in  dashed blue and 1.9 $\times$ 2 K in  dash-dotted green, all assuming 
$m_0 = 5$ meV, $\epsilon_{eg} =$ 11 meV and the zero chemical potential. 
Distortions are identical for the two cases of NH and IH.
}
\end{figure}

\begin{figure}
 \centering
\includegraphics[width=26em]{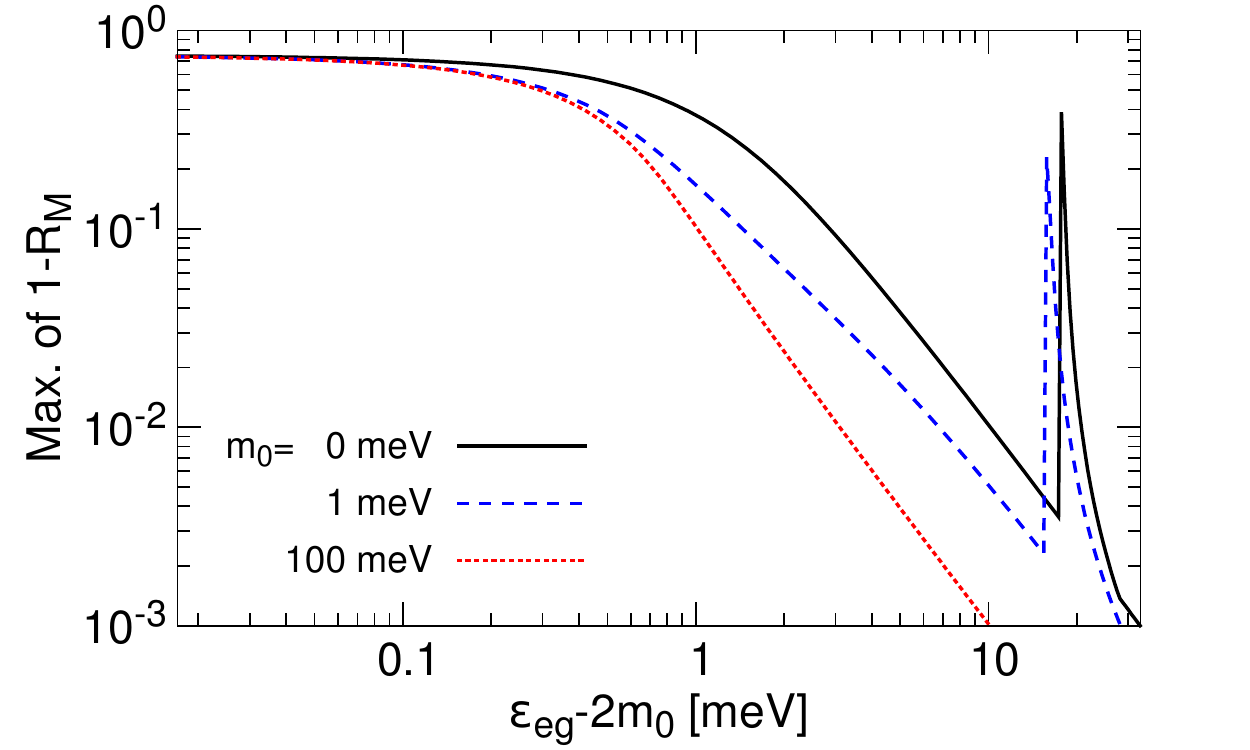}
 \caption{ \label {pauli 4ks 11-5-4 monopole}
Maxima of the spectral deviation $1-R_M(\omega)$ caused by 
          the Pauli blocking plotted against the difference between 
          level splitting and twice of the lightest neutrino mass, $\epsilon_{eg}-2m_0$.
         We show the cases of NH
          $m_0=0$ meV (solid black),  NH 1meV (dashed blue)
          and NH 100 meV (dotted red), assuming  the zero chemical 
          potential.}
\end{figure}

\begin{figure}
 \centering
 \includegraphics[width=26em]{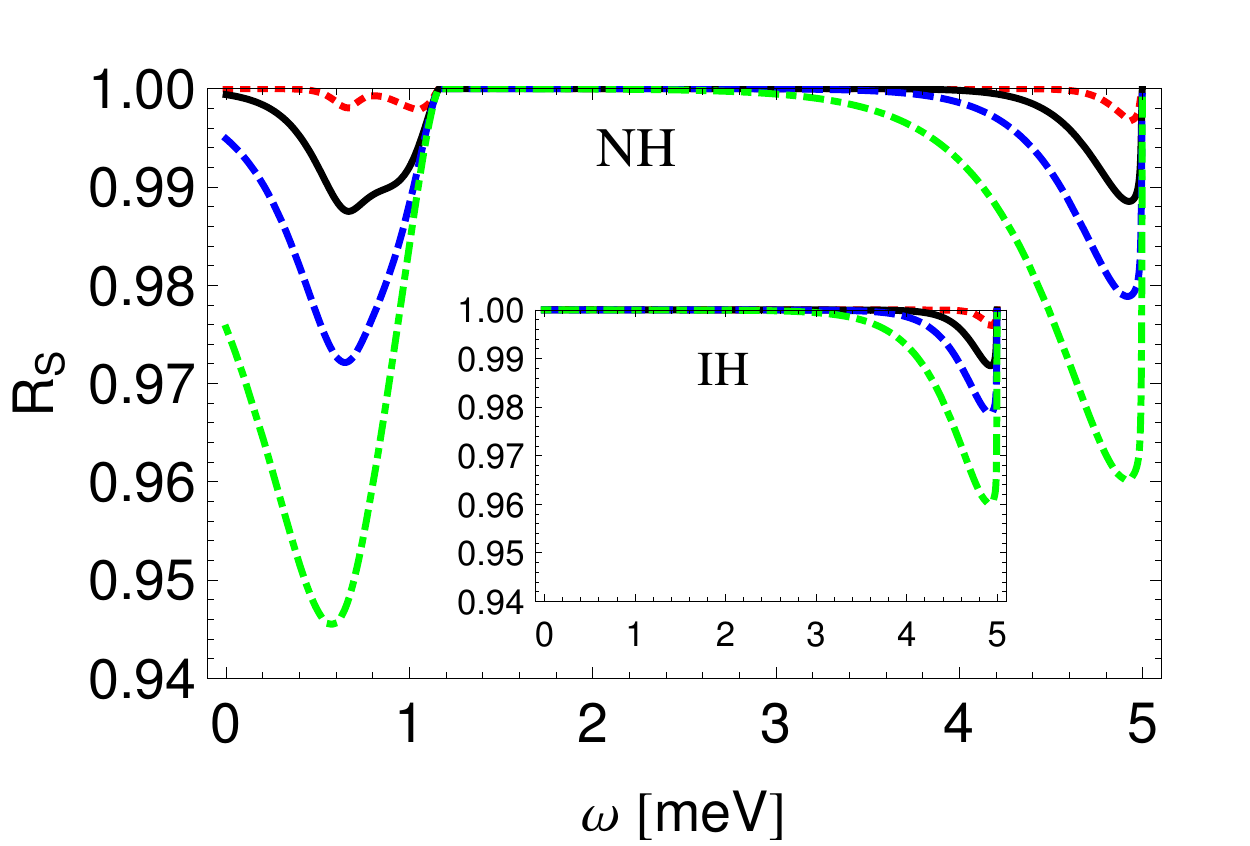}
   \caption{Spectral distortion $R_S(\omega)$ caused by the Pauli blocking.
$T_{\nu}= 1.9/2$ K in  dotted red, 1.9 K in solid black,
2.7 K in dashed blue and 1.9 $\times$ 2 K in  dash-dotted green,
all assuming 
$m_0 = 0.1$meV, $\epsilon_{eg} =$ 10 meV and  the zero chemical potential.
IH case is shown in the inset for comparison with NH case.
}
   \label  {pauli 4ks 10 spin}
\end{figure}

\begin{figure}[htbp]
 \centering
 \includegraphics[width=26em]{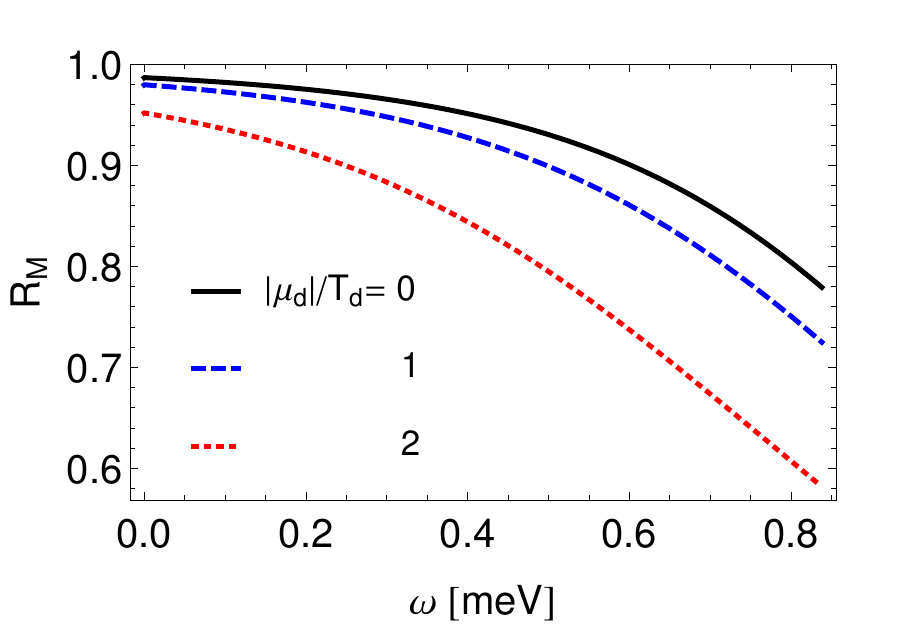}
 \caption{Spectrum distortion $R_M(\omega)$ for magnitudes
  of neutrino degeneracy $|\mu_d|/T_\nu=0$ meV in solid black,
1 in dashed blue, and 2 in dotted red. The lightest 
  neutrino mass $m_0=$ 0 meV.
$\epsilon_{eg} = 10 T_{\nu} \sim 1.7 $ meV chosen.
}
   \label {pauli monopole merelated}
\end{figure}

Calculated theoretical values of the spectral distortion
are shown in Fig(\ref {pauli 4ks 11}), Fig(\ref{pauli 4ks 11-5-4 monopole}) 
and Fig(\ref{pauli monopole merelated}) for the nuclear monopole contribution
and in Fig(\ref {pauli 4ks 10 spin}) for the spin current contribution.
Effects of non-vanishing CPV phases 
that appear in the weight factor of pair emission are small,  hence
for simplicity we assume the vanishing CPV phase in the following analysis.
Main results shown in Fig(\ref {pauli 4ks 11})
and Fig(\ref{pauli monopole merelated}), but neither in Fig(\ref{pauli 4ks 11-5-4 monopole}) 
nor in Fig(\ref {pauli 4ks 10 spin}),
are insensitive to which of
the neutrino mass hierarchical patterns, the normal or the inverted
hierarchy (NH, IH), is adopted, and results for these two cases are
identical.

The Pauli blocking effect becomes the largest in
the threshold region of neutrino pair emission of
smallest mass $m_0$.
In Fig(\ref  {pauli 4ks 11}) $\sim$ Fig(\ref{pauli monopole merelated})
we take hypothetical atoms of
excitation energy in the range, 0.1 $\sim$ 100 meV, 
and show the
Pauli blocking effect given by the rate ratio $R_A(\omega)$.
The difference between distortions of 1.9 K and 2.7 K,
of a crucial  importance to cosmology,
may reach 10 per cent level for
appropriate combination of $m_0$ and $\epsilon_{eg}$,
as in Fig(\ref {pauli 4ks 11}).

Study of relic neutrino detection
becomes more practical after RENP process  is discovered 
for a definite target atom and
a range of smallest neutrino mass is identified.
Anticipating an approximate $m_0$ determination already achieved, 
we present in Fig(\ref {pauli 4ks 11-5-4 monopole}) the maximal
spectral distortion
assuming a special relation between  $m_0$ and the atomic level spacing.
The peak structure for $m_0$ values of 0, 1 meV observed  in Fig(\ref {pauli 4ks 11-5-4 monopole}),
which shows a large distortion, 
is due to the second threshold $\omega_{22}$ of the next lightest
neutrino pair of mass $\sim 10$ meV.

At the zero momentum limit of $p=0$, $ 1-f_i \sim 1/2$ with the vanishing
chemical potential,
and the effect of Pauli blocking  becomes the largest.
The reason the largest distortion of $3/4$ is not realized in RENP unlike the
case of inverse process \cite{my 07 comment}
is that at thresholds $\omega_{ij}$ neutrinos cannot carry the zero momentum
and only a partial blocking occurs,
since the half energy $\sim \epsilon_{eg}/2$ is shared
by two neutrinos.

The absolute value of RENP spectral rates depends linearly on  
a time varying dynamical factor $\eta_{\omega}(t)$, which 
is the product of medium polarization and the stored field
energy in dimensionless units and may be
calculated by solving the master equation of coherence evolution 
\cite{renp overview}.
The nuclear monopole contribution \cite{nuclear monopole} gives the
largest rate  of order 50 events/second  $\times \eta_{\omega}(t)$ 
at its maximum for Xe atomic de-excitation of $^3P_1 (\sim 8.4$ eV)
for a gas target number density $7 \times 10^{19}$cm$^{-3}$ and 
a target volume $10^2$cm$^3$.
The large rate for a heavy atom $\propto Q_w^2 Z^{8/3}$ arises,
since the monopole charge
is proportional to $Q_w = N-0.044 Z$, with $N, Z$ the neutron number and
the proton number of nucleus. 
The rate is further enhanced by a large Coulomb interaction.
Dependence on atomic parameters is more complicated, but
very roughly the rate scales as $\propto $ the level spacing 
$\epsilon_{eg} \times$ relevant E1 dipole strength squared.
Although the rate near the threshold is suppressed,
it rapidly increases towards a maximum value at higher photon energies.

Distortions of 10 \% level or more seen in Fig(\ref{pauli 4ks 11})
and Fig(\ref{pauli 4ks 11-5-4 monopole}) for the nuclear monopole contribution
are experimentally encouraging.
The distortion of the photon spectrum in the spin current contribution
has an interesting second structure as shown in Fig(\ref {pauli 4ks 10 spin})
due to the second threshold of the neutrino pair of
smallest and the next smallest masses.
This second structure is present only in the NH case, and absent in the IH case
which gives identical results for the right side of structure to the NH case.
Rates are however much smaller than the monopole case \cite{nuclear monopole}.

The distorted spectrum for  a finite value of
the chemical potential has been calculated,
as illustrated in Fig(\ref{pauli monopole merelated}) \cite{massive majorana}.
For a choice of small $m_0$ effects of the finite chemical potential
may be non-negligible.

The prospect of relic neutrino detection is closely tied to
a success of neutrino mass spectroscopy using RENP.
We shall briefly describe the present status of our project and 
necessary investigation towards the final goal.

RENP uses the macro-coherence concept \cite{yst pra}  which
gives  the dependence $\propto n^3 V$ with
$n$ the target number density and $V$ the target volume, and the phase matching condition or 
the momentum
conservation among three light particles $\gamma, \nu_i, \bar{\nu}_j$.
The macro-coherence works when more than two light particles
are emitted in the final state, giving an important difference from
super-radiance that restricts the coherent region to wavelength squared \cite{sr}.
Recently we succeeded in experimentally observing the macro-coherent 
two-photon emission of a QED process, called Paired Super-Radiance (PSR)
\cite{psr observation}.
This indicates the enhancement mechanism ($> 10^{15}$ in rate) of
macro-coherence. 
A similar, but larger macro-coherence should also work in RENP.

Moreover, the macro-coherent PSR may be used for
development of the macro-coherence of RENP,
which ultimately leads to formation of static object
called soliton-condensate \cite{psr soiton}.
This is the remnant state of large stored light field coupled to
macroscopic medium polarization
after the termination of PSR related activity,
giving a stationary value of the dynamical factor $\eta_{\omega}$
as large as $10^{-3}$ or even more.
In the target state of soliton-condensates two-photon QED backgrounds are
exponentially suppressed, thus enhancing the signal to the background ratio.

Initial states of PSR and RENP processes have however different parities,
and one needs a switching mechanism between two different parities.
One of the ideas for this is the use of external electric field
to mix different parity states.
Experimental study of targets in solid environments is important to 
further narrow down a practical way of RENP experiment.

Finally, 
we mention the directional variation of distorted spectrum.
This is caused by the earth motion relative to
the 2.7 K microwave isotropic distribution, giving 
an effective momentum change of a dipole form in the neutrino
distribution function of order given by its velocity
 $v/c = O(10^{-3})$.
This effect should help identify relic RENP events
near the thresholds.

In summary,
neutrino mass spectroscopy using RENP
may become a sensitive tool to
explore the early cosmic epoch at
the decoupling of the electron neutrino.
We proposed to use distortion of the photon
energy spectrum caused by the Pauli blocking
of ambient relic neutrinos.
The sensitivity to the background temperature measurement
depends on the unknown mass value of lightest neutrino,
in relation to the level spacing of excitation.
The spectrum distortion may become large,
more than  significant  10\% level along with this order of temperature
distinction of 1.9 and 2.7 K.
A small level spacing thus favored may be provided by fine structure
splitting of atoms or in molecular rotational transitions.

\vspace{0.5cm}
We should like to thank
K. Inoue and M. Yamaguchi at Tohoku University for 
valuable discussions.
This research was partially supported by Grant-in-Aid for Scientific
Research on Innovative Areas "Extreme quantum world opened up by atoms"
(21104002) from the Ministry of Education, Culture, Sports, Science, 
and Technology, and JSPS KAKENHI Grant Number 25400257.


\begin{thebibliography}{99}
\bibitem{my 07 comment}
In \cite{my 07} the inverse process of
$\gamma + |g \rangle \rightarrow \nu_i \bar{\nu}_j + |e \rangle$ 
to RENP was considered, taking into account 
the Pauli blocking effect.
The process is  more difficult to experimentally detect than RENP,
even if it can be macro-coherently amplified. But
the Pauli blocking effect is larger and reaches the largest reduction factor 
$3/4$ in the range $m_0 \leq $ 1meV even for larger level spacing $\epsilon_{eg}$.
\bibitem{chemical poential}
One might consider that the chemical potential
is defined only in the case of Dirac neutrino
in which the particle number is well defined.
But weak interaction of Majorana neutrino
prior to decoupling  distinguishes
two different helicity states, which
may effectively be regarded as particle
and antiparticle as in the Dirac case.
Only the gravitational clustering that occurs
much later mixes two helicity states,
thereby making the chemical potential vanishing.
Since we do not need to consider the effect
of gravitational clustering, even the Majorana
case may be endowed with the chemical potential.
\bibitem{dirac}
We have computed the Dirac neutrino case in
a few examples.
In the case of Fig(\ref{pauli 4ks 11}) there
is no appreciable difference between the Majorana
and the Dirac cases.
\bibitem{massive majorana}
In Fig(\ref{pauli monopole merelated})
the case of $m_0=0$ meV (actually meaning $m_0 \ll \epsilon_{eg}/2$) is treated, because 
cases of finite $m_0$ values give less distortions 
for finite chemical potentials.
Moreover, one needs a formula of neutrino pair
production rates for helicity decomposed neutrino pairs,
requiring a more complicated formula than the
weighted sum of eq.(\ref {threshold kinematical factor}).
\end{thebibliography}
\end{document}